\documentstyle[amssymb,preprint,eqsecnum,aps,prb]{revtex}

\begin{document}
\title{Exchange coupling induced antiferromagnetic-ferromagnetic transition in $%
Pr_{0.5}Ca_{0.5}MnO_3/La_{0.5}Ca_{0.5}MnO_3$ superlattices.}
\author{P. Padhan and W. Prellier\thanks{%
prellier@ensicaen.fr}}
\address{Laboratoire CRISMAT, CNRS\ UMR 6508, ENSICAEN, 6 Bd du Mar\'{e}chal Juin,\\
F-14050 Caen Cedex, FRANCE.}
\date{\today}
\maketitle

\begin{abstract}
Superlattices built from two antiferromagnetic (AFM) charge/orbital order
compounds, $Pr_{0.5}Ca_{0.5}MnO_3$ and $La_{0.5}Ca_{0.5}MnO_3$, have been
studied as the thickness of $La_{0.5}Ca_{0.5}MnO_3$ ($LCMO$) varied. High
structural quality thin films were obtained on $LaAlO_3$ substrates using
the pulsed laser deposition technique. An antiferromagnetic-to-ferromagnetic
transition, in addition to an enhancement of the coercivity, are observed as
the $LCMO$ layer thickness increases. The small shift in the origin of the
field-cooled hysteresis loop along the field axis indicates the presence of
ferromagnetic and antiferromagnetic phases in the superlattices. We
attribute these features to the AFM spin fluctuations at the $%
Pr_{0.5}Ca_{0.5}MnO_3/La_{0.5}Ca_{0.5}MnO_3$ interfaces resulting from the
strain effects.
\end{abstract}

\newpage

In multilayer structures based on transition metal compounds several
fascinating magnetic properties such as oscillatory exchange coupling\cite
{1,2,3}, exchange bias\cite{4,5} and enhanced coercivity\cite{6} has been
observed. These magnetic phenomena are the interplay of exchange coupling at
the interfaces of the heterostructures composed of ferromagnetic ($FM$) and
non-magnetic, either metallic or insulating materials. In these
heterostructures, the interfaces are rich in magnetic and structural
coordinations of the transition metal ions\cite{7,8} through the interaction
processes like direct exchange, superexchange and double exchange. The
increase in coercivity is commonly observed when ferromagnetic thin film
coupled through the antiferromagnetic ($AFM$) thin film. Several possible
mechanisms have been used to explain the increased coercivity found in $%
FM/AFM$ systems such as the instabilities in the antiferromagnet\cite{9,10}
and inhomogeneous magnetization reversal\cite{11,12}. Another manifestation
of exchange coupling is the interfacial ferromagnetism at the interfaces of
the heterostructures. Ueda {\it et. al.}\cite{13} have study the magnetic
properties of the superlattices consisting of antiferromagnetic layers of $%
LaCrO_3$ and $LaFeO_3$ grown on ($111$)-oriented $SrTiO_3$ show a
ferromagnetic behavior. The authors have explained the ferromagnetic
behavior due to the ferromagnetic coupling between $Fe^{3+}$ and $Cr^{3+}$.
Takahashi {\it et. al.}\cite{14} have studied the transport and magnetic
properties of the superlattices made up of $AFM$ $CaMnO_3$ and paramagnetic $%
CaRuO_3$ grown on ($001$) oriented $LaAlO_3$ ($LAO$) show a Curie
temperature ($T_C$) at $\sim $ $95$ $K$ and negative magnetoresistance below 
$T_C$. The authors have concluded that the ferromagnetic-like transition
with appreciable spin canting occurs only near the interface region due to
the electron transfer from the $CaRuO_3$ layer to the $CaMnO_3$ layer
through the interface. Looking at these examples, it is interesting to built
superlattices in order to obtain novel electronic properties.\ For this,
many types of oxides can be used and mixed valance manganites is one of
them. Moreover, the manganite compounds exhibit many fascinating electronic
properties like colossal magneto-resistance (CMR), charge/orbital ordering.
The latter property of charge ordering has been seen in mixed valence
manganites in particular, when the dopant concentration is close to the
commensurate value $x$ $=$ $0.5$ (like $Pr_{0.5}Ca_{0.5}MnO_3$ and $%
La_{0.5}Ca_{0.5}MnO_3$) in the reduced bandwidth systems\cite{15,16}. In
these systems the charge-ordering gap can be collapsed by the application of
magnetic field, electric field, high pressure, optical radiation and
electron irradiation \cite{CO} and this results in a metal-like transport
below the charge-order transition temperature.

Here, we have synthesized superlattices consisting of two antiferromagnetic
insulator materials, $Pr_{0.5}Ca_{0.5}MnO_3$ ($PCMO$) and $%
La_{0.5}Ca_{0.5}MnO_3$ ($LCMO$), on ($001$)-oriented $LaAlO_3$ ($LAO$, cubic
with $a_{LAO}=3.79$ $\AA $) to investigate new magnetic and electronic
properties and our results are reported in this article. The effect of
strain-induced spin canting on the magneto-electronic properties with
various $LCMO$ layer thickness are studied keeping the $PCMO$\ layer at a
fixed thickness.

The samples were grown using the multitarget pulsed laser deposition
technique at $720$ $^{\circ }C$ in an oxygen ambient of $300$ $mtorr$\cite
{17a}. The deposition rates (typically $\symbol{126}$ $0.38$ $\AA /pulse$)
of $PCMO$ and $LCMO$ were calibrated for each laser pulse of energy density $%
\symbol{126}3$ $J/cm^2$. After the deposition the chamber was filled to $400$
$torr$ of oxygen at a constant rate, and then the samples were slowly cool
down to room temperature at the rate of $20$ $^{\circ }C/\min $. The
superlattice structures were synthesized by repeating $15$ times the bilayer
comprising of $20$-($unit$ $cell$, $u.c.$) $PCMO$ and $n$-($u.c.$) $LCMO$,
with $n$ taking integer values from $1$ to $20$. In all superlattices, the
top and bottom layers are $20$ $u.c.$ thick $PCMO$. The samples were
characterized by magnetization ($M$) in addition to resistivity ($\rho $)
and x-ray diffraction (XRD). Magnetization measurements were performed at $%
10 $ $K$ with magnetic field along the [$100$] and [$001$] directions of $%
LAO $.

The superlattices consisting of alternate layers of $PCMO$ and $LCMO$ grown
on ($001$)-oriented $LAO$ show ($00l$) diffraction peaks of the
constituents, indicating the growth of an epitaxial pseudocubic phase with
the $c$-axis orientation. The $\theta -2\theta $ scan for three samples with
different spacer layer thickness is shown in Fig.1(a). These scans are
recorded around the ($002$) reflection of these pseudocubic perovskites. The
first order satellite peak of the sample with $n$ $=$ $4$ and $12$ on the
higher angle side of the ($002$) diffraction peak of the constituents falls
on the ($002$) reflection of the substrate. While it is close to the ($002$)
reflection of the substrate for the sample with $n$ $=$ $20$. As the $LCMO$
layer thickness increases, the presence of higher order strong satellite
peaks on either side of the ($002$) diffraction peak, clearly indicates the
formation of a new structure having a periodic chemical modulation of the
constituents. The full-width-at-half-maximum ($FWHM$) of the rocking curve
correlates the structural coherence length $\xi $ of the sample with the
relation $\xi $ $=$ $\frac{2\pi }{Q\text{ }.\text{ }FWHM}$ \cite{18}, where $%
Q$ ($\approx $ $\frac 1d$) is the scattering vector length and $FWHM$ is in
radians. The coherence length along the [$001$] direction of the substrate,
for various samples with different $LCMO$ layer thickness, is shown in the
Fig. 1(b). The value of $\xi $ is several times the total thickness of the
superlattices, indicating the coherency\cite{17a} and confirming the single
crystallinity of the samples seen in the XRD data.

The temperature-dependent magnetization $M(T)$ was measured in the presence
of $0.1$ $tesla$ magnetic field, oriented along the [$001$] direction of the
substrate (i.e. within the plane). The field-cooled ($FC$) magnetization of
the superlattice with $n$ $=$ $4$ (Fig. 2a) on heating from $10$ $K$,
decreases slowly up to $150$ $K$, remains constant in the temperature range
of $150$ $K$ to $230$ $K$ and then again decreases slowly up to $320$ $K$.
This feature is qualitatively similar to that of the $PCMO$, i.e. the
superlattice with $n=4$ displays an AFM\ behavior\cite{17b}. As the $LCMO$
layer thickness increases up to $8$ $u.c.$ (Fig. 2b), the $FC$ magnetization
on heating from $10$ $K$, decreases slowly up to $60$ $K$, then it drops
rapidly till $170$ $K$. Above $170$ $K$, it decreases again slowly up to $%
320 $ $K$. The AFM behavior observed in the sample with $n=4$ is almost
suppressed in the sample with $n=8$. This AFM\ feature is completely
suppressed for superlattices with $n$ $\geq $ $10$ and the sample becomes
FM. As an example, the temperature-dependent magnetization for $n=12$ is
shown in Fig.2(c). The magnetization decreases very slowly above $10$ $K$ up
to $100$ $K$, above this temperature magnetization drops rapidly till $250$ $%
K$ and then decreases slowly up to $320$ $K$. This temperature-dependent
magnetization measured in spin equilibrium configuration (field-cooled)
correlates with the stronger ferromagnetic interaction at the interface.
Fig.2(c) displays, for $n=20$, the magnetization measured in spin
non-equilibrium (zero-field-cooled) and spin equilibrium configurations.
This figure shows a large difference between both configurations below $100$ 
$K$. This indicates the presence of an inhomogeneous nature of the spin
orientations at the interfaces as well as in the bulk, due to spin canting
or spin order. The increase in $LCMO$ layer thickness in the fixed $PCMO$
layer thickness based multilayers, clearly shows an
antiferromagnetic-to-ferromagnetic transition, which is confirmed by the
field-dependant magnetization described hereafter (see fig.3). Surprisingly,
for the $FM$\ samples (i.e. with $n\geq $ $10$), the Curie temperature of
the superlattices does not change significantly ($226$ $K$ and $229$ $K$ for 
$n$ $=$ $12$ and $n=20$, respectively) with the $LCMO$ layer thickness.

The enhancement of $FM$ is also observed in the field-dependent
magnetization $M$($H$) of the superlattices with the increase in magnetic
moments as the $LCMO$\ thickness increases. This is illustrated in the
zero-field-cooled ($ZFC$) $M$($H$) at $10$ $K$, recorded with a magnetic
field oriented along the [$100$] and [$001$] directions of the substrate,
for various samples ($n=4$, $8$, $12$ and $20$), shown in the Fig. 3. When
looking in details to the graph, we observed that the superlattice with $4$ $%
u.c.$ thick $LCMO$ layer shows $\approx $ $0.02$ $tesla$ coercive field ($%
H_C $) for both orientations of the magnetic field (Fig.3a). It also shows a
small anisotropy, while the magnetization increases gradually with the
increase in either in-plane or out-of-plane magnetic field. A qualitatively
similar hysteresis loop (Fig.3b, c and d), but with a higher value of the
coercive field, is observed for the sample with higher thickness of $LCMO$
layer ($n=8$, $12$ and $20$). Moreover, for the samples with $n$ $>$ $6$,
the in-plane coercive field is smaller than the out-of-plane coercive field.
This difference is clearly seen in the Fig.4(a) where the in-plane and
out-of-plane coercive fields for various samples are plotted. The $H_C$
increases with the increase in $LCMO$ spacer layer thickness and saturates
for the sample with $n$ $>$ $10$. From this figure, it is observed that the
anisotropy in $H_C$ appears for sample with $n$ $>$ $6$. This anisotropy
increases up to $n$ $\approx $ $12$ and remains the same for higher value of 
$n$. Although a relatively small increase in $H_C$ has been observed in the
superlattices with $n$ $\leqslant $ $6$ compared to its constituents ($LCMO$%
\ and $PCMO$). Nevertheless, the exchange coupling at the interfaces is
strongly enhanced $H_C$ for superlattices with $n$ $>$ $6$. Since the
magnetic interactions between the $Mn$ ions in the bulk $PCMO$ or $LCMO$ do
not lead to the enhancement of $H_C$ , the origin of the enhancement must be
from the exchange interaction between $PCMO$ and $LCMO$ at the interfaces.
The fact that such features are strongly dependant on the stacking of the
superlattices and viewing some recent results\cite{19,19a} reinforce this
statement.

For the ideal antiferromagnetic state of the constituents, the magnetization
of $PCMO/LCMO$ should be independent of the magnetic field. The gradual
increase in magnetization for both orientations of the magnetic field in the
hysteresis loop (Fig.3), indicates that the $AFM$ sublattices contribute to
the coupling energy at the interfaces, when the difference in the
orientation of its two magnetization sublattices deviates from $180{%
{}^{\circ }}$. The origin of the reorientation of the spins of the $AFM$
sublattices could be due to the $3D$ coordinations of different $A$-site
ions and/or the inhomogeneous magnetic phases. This will induce an extra
interfacial anisotropy, and hence the anisotropy in the coercivity. The
fluctuations of the $AFM$ spin at the interfaces enhance the coercive field
with the increase in $LCMO$ layer thickness\cite{19}. This effect is also
realized in the net magnetization of the superlattices. The net
magnetization of the superlattices at $1$ $tesla$ magnetic field with two
orientations, at $10$ $K$, for samples with various $LCMO$ layer thicknesses
are shown in the Fig. 4(b). As the $LCMO$ layer thickness increases from 1
u.c. to 10 u.c. the magnetization in $M$($H$), recorded at $10$ $K$, under $1
$ $tesla$ magnetic field increases two times, and for higher value of LCMO
layer thickness the magnetization increases to a negligibly small value. To
explain these observations we consider coherency and intrinsic
inhomogeneities of the constituents. The presence of two different ionic
size elements at the A-site in $PCMO$ and LCMO leads to an intrinsic
inhomogeneities\cite{19b}. However, it is important at the interfaces due to
the presence of La, Pr and Ca. This introduction of of an inherent or
quenched disordered in the system results in a low-temperature regime that
consists of ferromagnetically or antiferromagnetically ordered phases
(inhomogeneous magnetic phases)\cite{19d} with randomly oriented order
parameters. The presence of inhomogeneous magnetic phases in the bulk leads
to three possible local magnetic coordination (AFM-AFM, AFM-FM and FM-FM) at
the interfaces. The increase in $LCMO$ layer thickness, i.e. the relaxation
of strain, varies the strength of the exchange coupling at $PCMO/LCMO$ the
interfaces. As the $LCMO$ layer thickness increase from 1 u.c. to 9 u.c. the
increase in magnetization is due to the spin reorientation of the $AFM$
sublattice at the interfaces. However, the relaxation of strain also induces
its bulk-like properties in the $LCMO$ layer. For ideal antiferromagnetic $%
LCMO$ the magnetization of the sample with n $\geq $ 10 should saturate. But
the non-significant increase in magnetization for sample with n $\geq $ 10
could be due to the presence of inhomogeneous magnetic phases with the
increase in microscopic to mesoscopic $FM$ order parameters in $LCMO$.

We have performed more measurements to confirm the $AFM$\ spin fluctuations
at the $PCMO/LCMO$ interface. In fact, the magnetic interactions across the
interfaces between a ferromagnetic spin system and an antiferromagnetic spin
system are generally known as exchange coupling, with phenomenological
features such as enhancement of coercive field $H_C$ and a shifted
hysteresis loop in the direction of the magnetic field\cite{4,5}. It is
usually observed on cooling the $FM/AFM$ system below the Curie temperature
of the $FM$ through the Neel temperature $T_N$ of the $AFM$ in presence of
magnetic field. We have used this formalism to verify whether the
fluctuations of the $AFM$ spin at the interfaces leads to the inhomogeneous
magnetic phases in this system. The $ZFC$ and $FC$ hysteresis loops of the
sample with $n$ $=$ $20$ at $10$ $K$ are shown in the Fig. 4(c). Though the
constituents materials are antiferromagnetic, as the sample is cooled below
room temperature in presence of $2$ $tesla$ magnetic field, the origin of
the hysteresis loop is shifted towards the negative field axis. This
confirms the presence of magnetic inhomogeneity in the samples.

We now tried to correlate these measurement with the transport as well as
the structure of the samples. Thus, we have also analyzed the structure and
transport properties of these samples as a function of $LCMO$ layer
thickness. In oxide thin films, it is well known that the structural and
transport properties are strongly dependent on the strains imposed by the
substrate. This is particularly true for $PCMO$ and $LCMO$ thin films as
previously observed in similar films\cite{20,21}. The lattice parameter of
bulk $PCMO$ ($a_{PCMO}$ $=$ $3.802$ $\,\AA $) and $LCMO$ ($a_{LCMO}$ $=$ $%
3.83$ $\,\AA $) is larger than $a_{LAO}$ with a lattice mismatch $+$ $0.3$ $%
\%$ and $+$ $1.05$ $\%$. Indeed, the epitaxial growth of $PCMO$ on $LAO$
provides in-plane compressive stress on $PCMO$. Similar kind of stress is
also expected at the interfaces for the epitaxial growth of $LCMO$ on $PCMO$
and such difference might affect the physical properties. In the
superlattices, the out-of-plane lattice parameter `$c$' increases with the
increase in spacer layer thickness and saturates for the sample with $n$ $>$ 
$10$ (Fig. 5a). The $c$-axis lattice parameter of the superlattice with $n$ $%
=$ $1$ increases to $\approx $ $0.3$ $\%$ as $n$ increases to $20$. This
change is equal to the lattice mismatch between $LAO$ and $PCMO$. Thus, we
conclude that the substrate-induced stress plays an important role in the
structure of the superlattices similarly to any manganite films\cite{21}.
However, the relaxation does not change qualitative behavior of temperature
dependent resistivity, but increase the conducting path. This leads to a
lower in the resistivity of the sample with the increase in the $LCMO$ layer
thickness. As the sample is cooled below room temperature down to 100 K, it
gains three orders resistivity. This significant change in resistivity with
temperature does not show remarkable variation in the $LCMO$ thickness
dependence resistivity curve at different temperature (100 K and 300 K in
Fig. 5b). As the resistivity of all samples with various $LCMO$ layer
thickness is very high below 100 K, it prohibits to compare the $LCMO$
thickness dependence resistivity below 100 K. Thus we present the change in
the magnetoresistance ($MR$ $=$ [$\rho $($0$) $-$ $\rho $($H$)]/$\rho $($H$%
)) at 100 K (Fig.5c) as a function of the $LCMO$\ thickness. This notation
for the $MR$ is used for the better resolution at the higher $LCMO$ layer
thickness. The crossover region from strained to strain-relaxed state with $%
LCMO$ layer thickness appears in the same region (close to n = 8) as those
observed in the variation of the coercive field (Fig. 4a, at $10$ $K$),
magnetization (Fig. 4b, at $10$ $K$), resistivity and magnetoresistance with
the $LCMO$ layer thickness, indicate that the charge-spin coupling is
correlated with the structure. This also suggests that both the
crystallographic and/or magnetic reconstructions and relaxation are
responsible for the physical properties of this system.

In conclusion, the superlattices composed of $Pr_{0.5}Ca_{0.5}MnO_3$ and $%
La_{0.5}Ca_{0.5}MnO_3$ compounds were grown on $(100)-LaAlO_3$ using pulsed
laser ablation. The fixed $PCMO$ layer based $PCMO/LCMO$ superlattices show
an antiferromagnetic-to-ferromagnetic transition with the increase in the $%
LCMO$ layer thickness. The coercive field, magnetization at $1$ $tesla$, $c$%
-axis lattice parameter, resistivity and magnetoresistance show a cross over
to their saturation values for the same $LCMO$ layer thickness. We attribute
these correlations to the crystallographic and/or magnetic reconstructions
and relaxations at the $PCMO/LCMO$\ interfaces. The coercive field is
anisotropic to the orientations of the magnetic field due to the magnetic
inhomogeneity along the out-of-plane direction of the substrate. An
enhancement of coercivity is observed in the superlattices with $n$ $>$ $6$.
We have interpreted this enhancement as the $AFM$ spin fluctuations at the
interfaces. The presence of magnetic inhomogeneity is also confirmed from
the ZFC and FC hysteresis loop of the superlattices. The transport behavior
of the superlattices are similar to that of its constituents (i.e.
insulating) but the increase in $LCMO$ layer thickness induced lower
resistive conduction path. This study confirms the importance of the
interfaces in superlattices that can be use to control novel physical
properties in oxide materials.

We greatly acknowledged financial support of Centre Franco-Indien pour la
Promotion de la Recherche Avancee/Indo-French Centre for the Promotion of
Advance Research (CEFIPRA/IFCPAR) under Project N${{}^{\circ }}$%
2808-1.

\bigskip

Fig.1(a): Reflected intensity of $\Theta -2\Theta $ scan recorded around the 
$002$ reflection of $LAO$ for various superlattices. The satellite peaks of
several orders (from $-$ $3$ to $+$ $3$) around the main peak (order $0$)
are indicated by arrows. $n$ reprensents the number of LCMO\ layer in the
PCMO/LCMO\ superlattice. (b) Evolution of the coherence length of the
superlattices with different $LCMO$ layer thickness. The solid line is a
guide to the eyes.

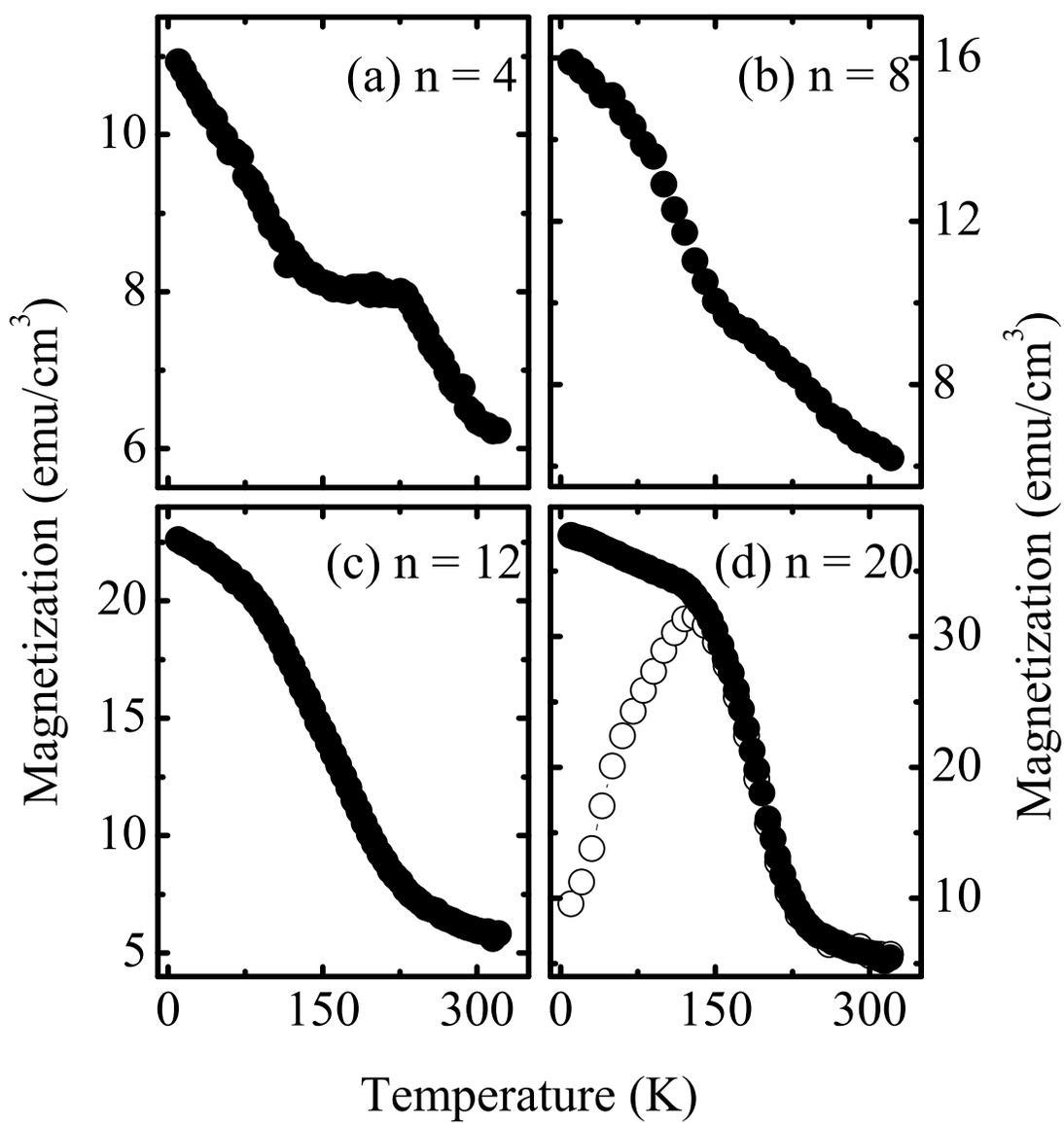
Fig.2: Field-cooled temperature dependent magnetization (filled circle) at $%
10$ $K$ at $0.1$ $tesla$ out-of-plane magnetic field of various
superlattices(panel a: $n=4$, panel b: $n=8$ and panel c: $n=12$). Panel d
shows zero-field-cooled temperature dependent magnetization (open circle)
and field-cooled temperature dependent magnetization (filled circle) of the
superlattice with $n$ $=$ $20$ at $10$ $K$ at $0.1$ $tesla$ out-of-plane
magnetic field.

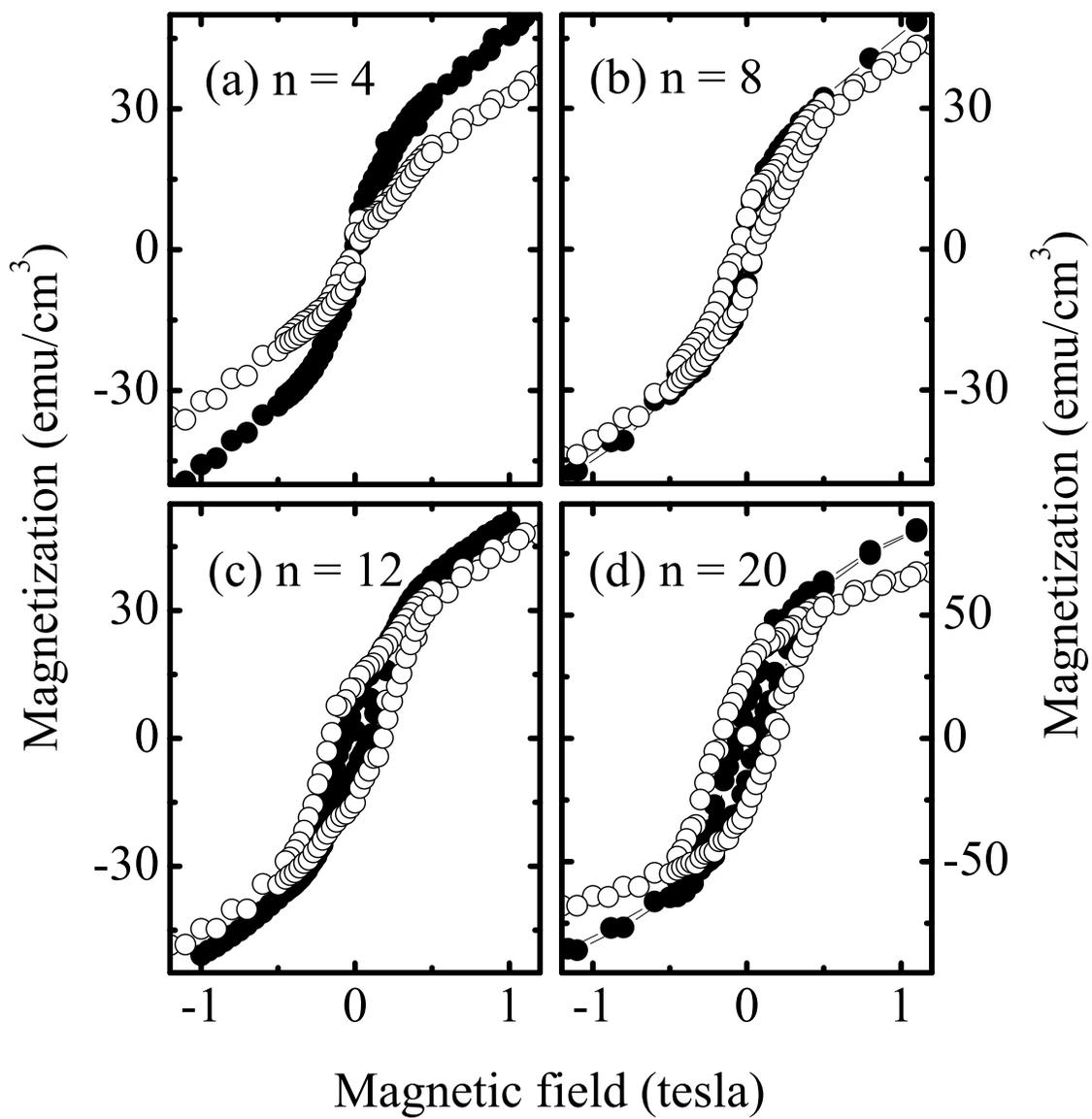
Fig.3: Zero-field-cooled magnetic field dependent magnetization along
in-plane (filled circle) and out-of-plane (open circle) directions of the
superlattices with $n$ $=$ $4$, $8$, $12$ and $20$ at $10$ $K$.

Fig. 4(a): In-plane and out-of-plane coercive field at $10$ $K$ of the
superlattices with different $LCMO$ layer thicknesses. (b) $ZFC$
magnetization of various samples with different $LCMO$ layer thicknesses at $%
10$ $K$ at $1$ $tesla$ magnetic field. The solid lines are guides to the
eyes. (c) Zero-field-cooled and field-cooled ($2$ $tesla$) magnetic field
dependent magnetization along out-of-plane directions of the superlattices
with $n$ $=$ $20$ at $10$ $K$.

Fig. 5(a) (b) and (c) Evolution of the out-of-plane lattice parameter,
resistivity at $100$ $K$ and magnetoresistance at $100$ $K$ under $7$ $tesla$
applied magnetic field respectively of the superlattices for different $LCMO$
layer thicknesses. The solid lines are guides to the eyes.

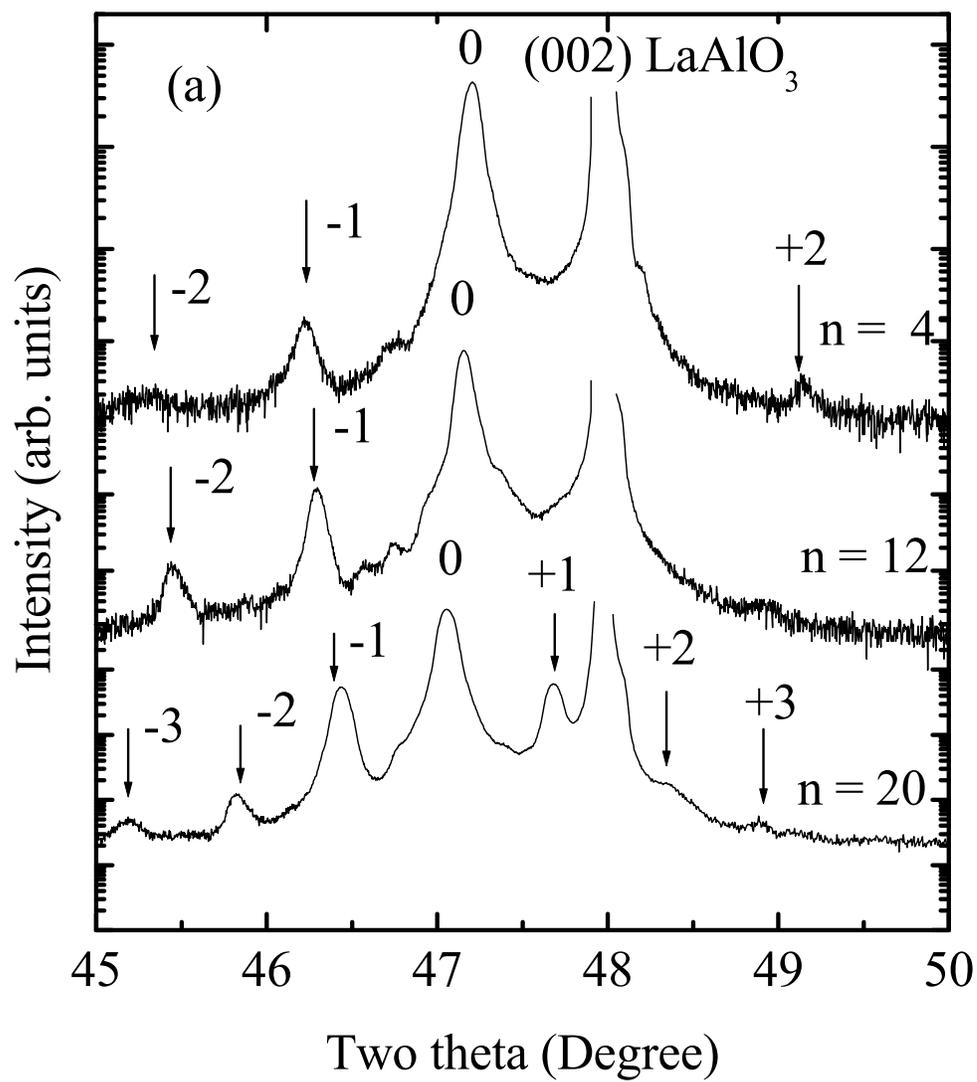
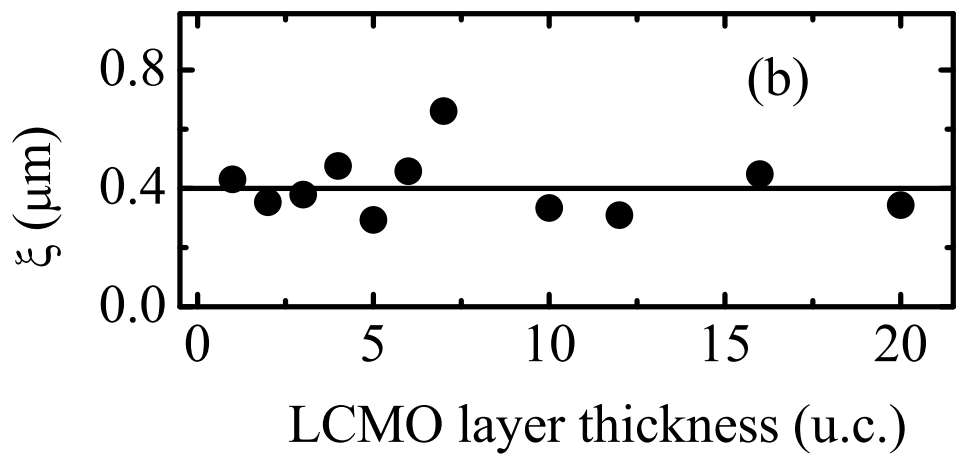
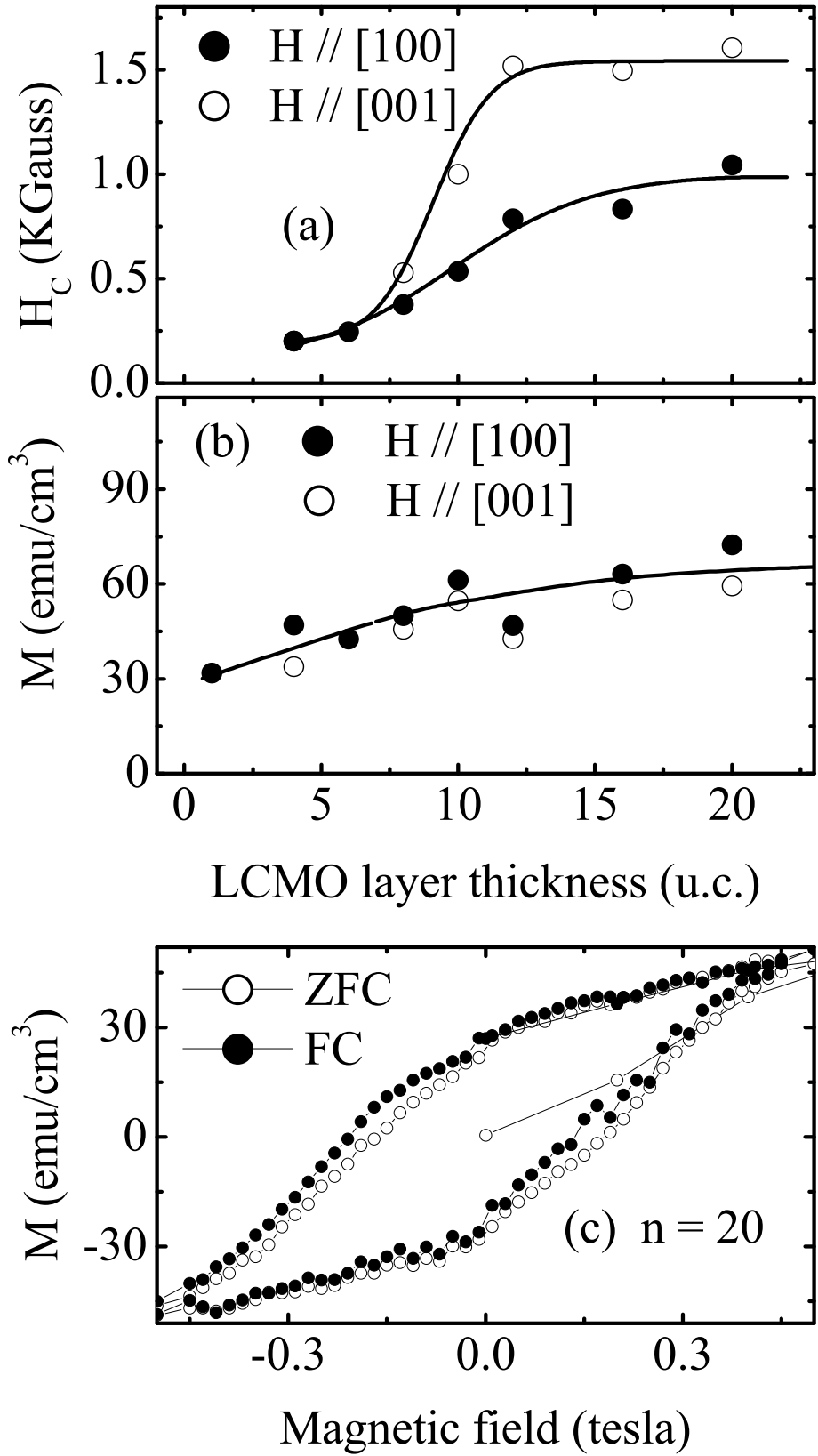
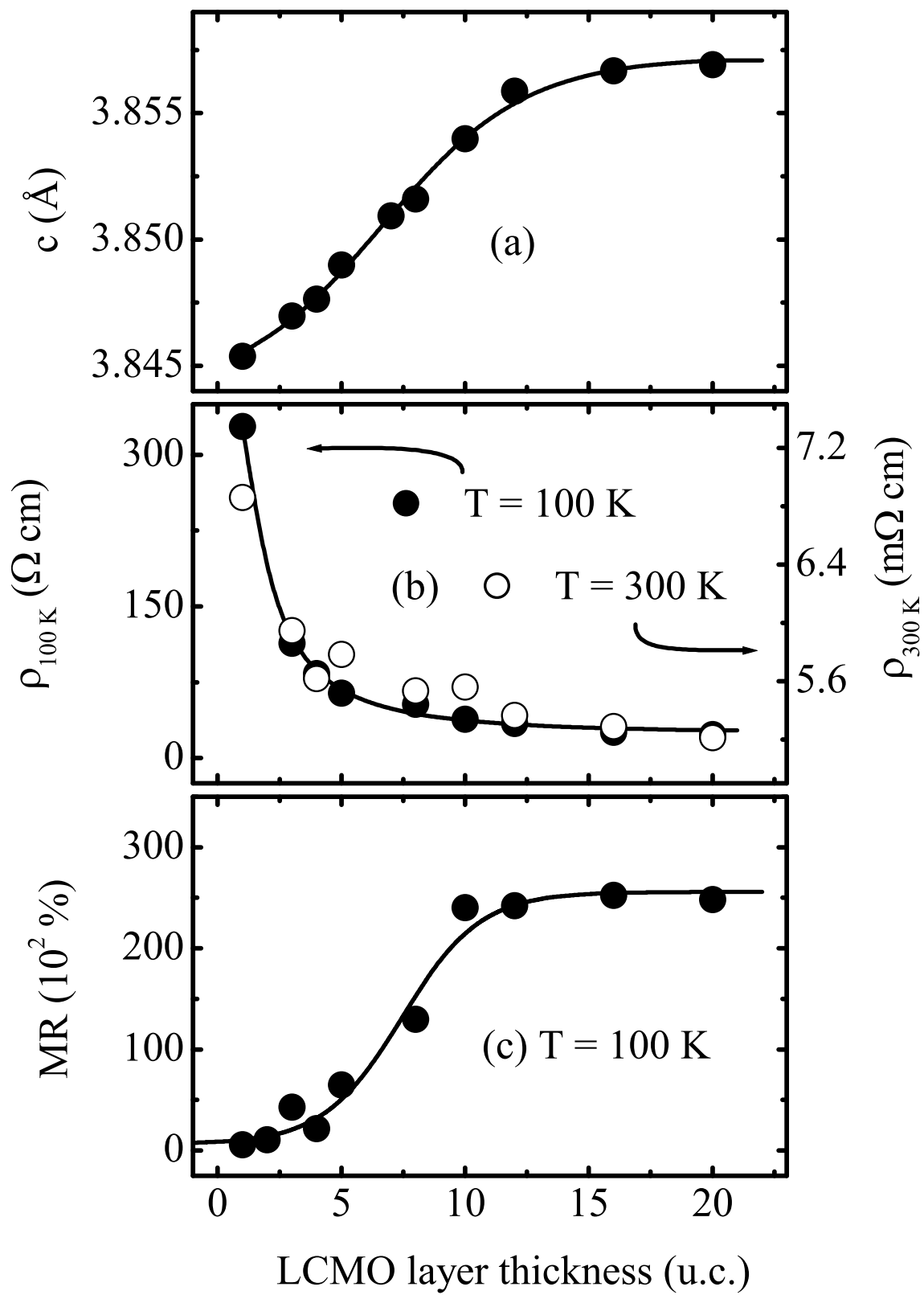
\end{document}